\documentclass[iop]{emulateapj}

\usepackage{multirow}

\newcommand{\xmm} {{\it XMM-Newton}}
\newcommand{\chandra} {{\it Chandra}}
\newcommand{\nustar} {{\it NuSTAR}}

\newcommand{\swift} {{\it Swift}}
\newcommand{\swiftxrt} {{\it Swift}/XRT}

\newcommand{\cmsq} {cm$^{-2}$}
\newcommand{\nh} {$N_{\rm{H}}$}
\newcommand{\lx} {$L_{\rm{X}}$}
\newcommand{\fx} {$F_{\rm{X}}$}
\newcommand{\chisq} {$\chi^2$}
\newcommand{\dchisq} {$\Delta\chi^2$}

\newcommand{\ergs}{\mbox{\thinspace erg\thinspace s$^{-1}$}}
\newcommand{\ergcms}{\mbox{\thinspace erg\thinspace cm$^{-2}$\thinspace s$^{-1}$}}

\newcommand{\mbh} {$M_{\rm BH}$}

\newcommand{\msol} {$M_{\odot}$}

\newcommand{\fscatt} {$f_{\rm scatt}$}

\shorttitle{.}
\shortauthors{Brightman et al.}

\begin{document}

\title{Spectral evolution of the ultraluminous X-ray sources M82 X-1 and X-2}

\author{Murray Brightman$^{1}$, Dominic J. Walton$^{2}$, Yanjun Xu$^{1}$, Hannah P. Earnshaw$^{1}$, Fiona, A. Harrison$^{1}$, Daniel Stern$^{3}$, Didier Barret$^{4,5}$}

\affil{$^{1}$Cahill Center for Astrophysics, California Institute of Technology, 1216 East California Boulevard, Pasadena, CA 91125, USA\\
$^{2}$Institute of Astronomy, Madingley Road, Cambridge CB3 OHA, UK\\
$^{3}$Jet Propulsion Laboratory, California Institute of Technology, Pasadena, CA 91109, USA\\
$^{4}$Universite de Toulouse, UPS-OMP, IRAP, Toulouse, France\\
$^{5}$CNRS, IRAP, 9 Av. colonel Roche, BP 44346, F-31028 Toulouse cedex 4, France\\
}

\begin{abstract}

M82 hosts two well-known Ultraluminous X-ray sources (ULXs). X-1, an intermediate-mass black hole (IMBH) candidate, and X-2, an ultraluminous X-ray pulsar (ULXP). Here we present a broadband X-ray spectral analysis of both sources based on ten observations made simultaneously with \chandra\ and \nustar. \chandra\ provides the high spatial resolution to resolve the crowded field in the 0.5--8 keV band, and \nustar\ provides the sensitive hard X-ray spectral data, extending the bandpass of our study above 10 keV. The observations, taken in the period 2015--2016, cover a period of flaring from X-1, allowing us to study the spectral evolution of this source with luminosity. During four of these observations, X-2 was found to be at a low flux level, allowing an unambiguous view of the emission from X-1. We find that the broadband X-ray emission from X-1 is consistent with that seen in other ULXs observed in detail with \nustar, with a spectrum that includes a broadened disk-like component and a high-energy tail. We find that the luminosity of the disk scales with inner disk temperature as $L{\propto}T^{-3/2}$ contrary to expectations of a standard accretion disk and previous results.  These findings rule out a thermal state for sub-Eddington accretion and therefore do not support M82 X-1 as an IMBH candidate. We also find evidence that the neutral column density of the material in the line of sight increases with \lx, perhaps due to an increased mass outflow with accretion rate. For X-2, we do not find any significant spectral evolution, but we find the spectral parameters of the phase-averaged broadband emission are consistent with the pulsed emission at the highest X-ray luminosities.

\end{abstract}


\section{Introduction}
Ultraluminous X-ray sources (ULXs), first discovered by the {\it Einstein X-ray observatory} \citep{giacconi79,fabbiano89}, are observed as bright sources of X-rays which appear to exceed the Eddington limit of the typical 10 \msol\ black holes found in our own Galaxy. ULXs are not coincidental with the nuclei of their host galaxies, so are not be powered by the supermassive black holes (SMBHs) that power active galactic nuclei (AGN), although $\sim$25\% of candidate ULXs are likely to be background AGN \citep[e.g.][]{walton11,sutton15,earnshaw19}. At first ULXs were promising intermediate mass black hole (IMBH) candidates, since larger black holes can radiate at higher luminosities due to the Eddington limit scaling with mass \citep[e.g.][]{colbert99,miller03}. But as more observations were made, the data did not appear consistent with this scenario, and instead, lower-mass black holes accreting at super-Eddington luminosities were favored for most sources \citep[e.g.][]{mizuno99}, with a few IMBH candidates still remaining \citep[e.g. ESO 243-49 HLX-1,][]{farrell09}.

M82 \citep[The `cigar galaxy', or NGC 3034,][]{watson84} hosts two well-known ULXs, that since are separated by only 5\arcsec\ on the sky, were only first resolved by \chandra\ \citep{matsumoto01}. The history of studies of M82 X-1 (CXOU J095550.2+694047), the brightest ULX in M82, embodies the above narrative. It has long been considered one of the best IMBH candidates because of its high luminosity, which can reach $\sim10^{41}$~\ergs\ \citep[e.g.][]{ptak99b, rephaeli02, kaaret06}; detection of low-frequency quasi-periodic oscillations (QPOs) in the power spectrum \citep[54 mHz,][]{strohmayer03, dewangan06, mucciarelli06}, indicative of a compact, unbeamed source; as well as twin-peaked QPOs at 3.3 and 5.1 Hz, which lead to a mass estimate of 400 \msol\ using scaling laws between the QPOs frequencies and mass used for stellar-mass black holes \citep{pasham14}. Additionally, \cite{feng10}  (hereafter F10) observed the source with \xmm\ and \chandra\ over the course of a flaring episode and fitted the spectra with the standard thin accretion disk model. They found that the luminosity of the disk, $L$, scaled with inner temperature as $L\propto T^4$ which is expected from a thin accretion disk with a constant inner radius. From this they inferred a black hole mass in the range $300-810$~\msol, assuming that the black hole is rapidly spinning in order to avoid extreme violations of the Eddington limit, therefore adding support to the IMBH scenario.

The luminosity argument no longer stands however, since the ULX NGC 5907 ULX1, which also reaches $\sim10^{41}$~\ergs\ \citep{sutton13,fuerst17}, and was once considered an IMBH candidate, was discovered to be powered by a neutron star with only 1--2 \msol\ \citep{israel17a} from the detection of coherent pulsations. The mass measurement from the twin QPOs is ambiguous too, since it is not known where these originate and it is not clear if the scaling relationship used extends to the IMBH range. In \cite{brightman16c}, hereafter B16, we presented a combined spectral analysis of X-1, also during a flaring episode, using simultaneous observations with \chandra, \nustar\ and \swift. With broader band data than \cite{feng10}, we found that the temperature profile as a function of disk radius ($T(r)\propto r^{-p}$) is significantly flatter than expected for a standard thin accretion disk as implied by F10, and instead characteristic of a slim disk that is expected at high accretion rates. Since only one observation was analyzed, the $L\propto T^4$ relationship could not be tested. Nevertheless, the mass estimates inferred from the inner disk radius for this model imply a stellar-remnant black hole (\mbh=$26^{+9}_{-6}$~\msol) when assuming zero spin, or an IMBH (\mbh=$125^{+45}_{-30}$~\msol) when assuming maximal spin. 

M82 X-2 (CXOM82 J095551.1+694045) is typically the second brightest X-ray sources in the galaxy, and was the first ultraluminous X-ray pulsar discovered \citep{bachetti14} using \nustar\ \citep{harrison13}. With only a 5\arcsec\ angular separation from X-1, studying the spectral properties of this source in detail has been limited to \chandra\ observations. We conducted a study of the spectral and temporal properties of M82 X-2 in \cite{brightman16} finding that the source's luminosity varies over two orders of magnitude over the range $10^{38}-10^{40}$ \ergs. Using timing analyses, we were able to isolate the pulsed emission from this source with \nustar, finding that it was well described by a cutoff power-law. In a follow up study, we found evidence that the variations in luminosity are modulated on a $\sim60$-day period, and that since the orbit of the neutron star and its companion is known to be 2.5 days, the $\sim60$-day period must be super-orbital in origin \citep{brightman19}. 

In this paper we present analysis of a systematic monitoring campaign on M82 by \chandra\ which took place in 2016. The primary goals of this campaign were to perform a temporal analysis of X-1 and X-2 to search for orbital and super-orbital modulations; to perform spectroscopic studies of the ULXs; and to study the nature of the other binary systems in M82. We report the temporal analysis in \cite{brightman19} and here we focus on the second objective, to present the most comprehensive X-ray spectral analysis of the two ULXs, M82 X-1 and X-2, to date. We combine simultaneous observations with \chandra\ to spatially resolve the two sources from each other and the other sources in M82, and simultaneous observations with \nustar\ to extend the spectral coverage up to 79 keV. We also present results from \swift\ monitoring observations of M82 which have been ongoing since 2012. Throughout this paper we assume a distance to M82 of 3.3 Mpc \citep{foley14} which is inferred from the lightcurve of SN2014J.

\section{Observations and Data reduction}
\label{sec_obs}

All \chandra\ and \nustar\ observations studied here were taken simultaneously, or quasi-simultaneously (within 24 hours of each other). Table \ref{table_obsdat} provides a description of the observational data. The following sections describe the individual observations and data reduction. 

\begin{table}
\centering
\caption{Observational data}
\label{table_obsdat}
\begin{center}
\begin{tabular}{l c c c c r c l}
\hline
Observatory	& Start date & Start time  & ObsID	&  Exposure	 \\
			&		&		(UT)	& 						& (s) 	 \\
\nustar\ &  2015-01-15 &    21:41:07 & 50002019002 &      31249 \\
\chandra\ &  2015-01-16 &    13:40:00 &       17578 &      10070 \\
\nustar\ &  2015-06-20 &    14:21:07 & 90101005002 &      37409 \\
\chandra\ &  2015-06-21 &    02:45:09 &       17678 &      10100 \\
\chandra\ &  2016-01-26 &    19:44:49 &       18062 &      25100 \\
\nustar\ &  2016-01-26 &    20:06:08 & 80202020002 &      36136 \\
\nustar\ &  2016-02-23 &    17:01:08 & 80202020004 &      31670 \\
\chandra\ &  2016-02-24 &    00:37:06 &       18063 &      25100 \\
\nustar\ &  2016-04-05 &    09:41:08 & 80202020006 &      30505 \\
\chandra\ &  2016-04-05 &    16:04:41 &       18064 &      25090 \\
\nustar\ &  2016-04-24 &    19:16:08 & 80202020008 &      40357 \\
\chandra\ &  2016-04-24 &    20:02:13 &       18068 &      25100 \\
\nustar\ &  2016-06-03 &    20:41:08 & 30202022002 &      39022 \\
\chandra\ &  2016-06-03 &    22:10:58 &       18069   &    25100 \\
\nustar\ &  2016-07-01 &    16:26:08 & 30202022004 &      47043 \\
\chandra\ &  2016-07-01 &    23:18:08 &       18067    &   26100 \\
\chandra\ &  2016-07-29 &    07:50:01 &       18065   &    25100 \\
\nustar\ &  2016-07-29 &    23:06:08 & 30202022008   &    42846 \\
\nustar\ &  2016-10-07 &    20:21:08 & 90202038002 &	45476 \\
\chandra\ &  2016-10-08 &    00:39:30 &       18070     &  23200 \\

 \hline
\end{tabular}
\end{center}
\end{table}

\subsection{Chandra}
Since the angular separation of X-1 and X-2 is only 5\arcsec, only \chandra\ \citep{weisskopf99} can spatially resolve the emission from these two sources. The majority of the \chandra\ data analyzed here were taken during 2016 (Cycle 17) as part of a Large Program aimed at systematic monitoring of binaries in M82. The program consisted of 12 individual observations taken at $\sim$monthly intervals, but only 8 having simultaneous \nustar\ observations were used in this work.  We additionally use two observations taken in 2015 which also have a simultaneous \nustar\ observation. Full details are listed in Table \ref{table_obsdat}. All 2016 observations were taken with ACIS-I at the optical axis using a 1/8th sub-array of pixels on chip I1 or I3, depending on the roll angle. The ULXs at the center of M82 were placed 3\arcmin.5 off-axis to smear out the PSF enough to reduce the effects of pile-up, but not so much as to cause significant blending of the PSFs. The sub-array of pixels was used to decrease the readout time of the detector to 0.4\,s, further reducing the effects of pile-up.

\begin{figure}
\begin{center}
\includegraphics[width=80mm]{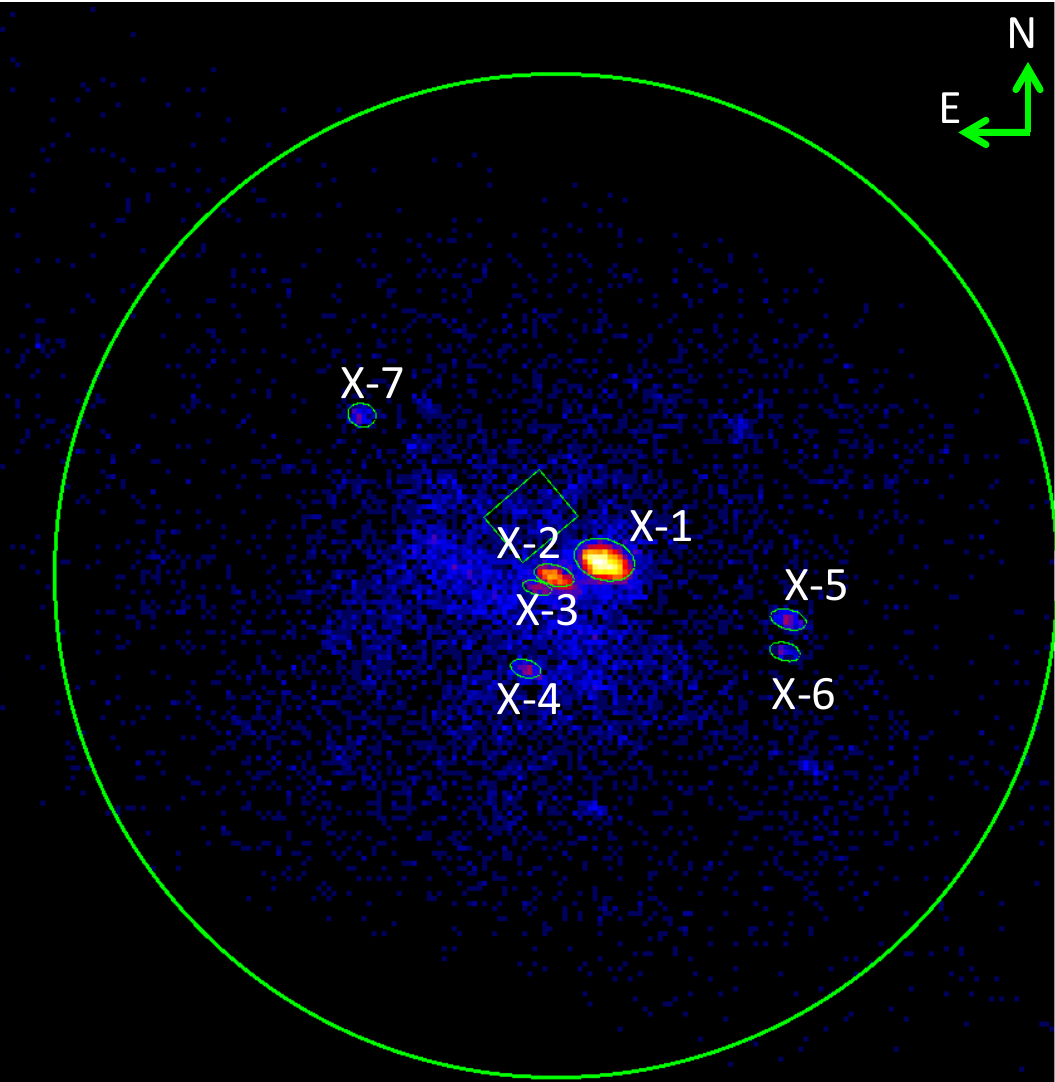}
\caption{\chandra\ image of M82 from obsID 17678 showing examples of the various extraction regions used,  including the background for \chandra\ analysis shown as a small rectangular region. Small ellipses are used for X-1 and X-2, whereas a large circle with a radius of 49\arcsec\ is used to extract events from the rest of the galaxy and the \nustar\ and \swift/XRT events. The brightest point-sources within this region are labelled, however X-1 and X-2 dominate. North is up, east is left, indicated by the arrows in the upper right corner, which are 10\arcsec\ long. }
\label{fig_chandra_img}
\end{center}
\end{figure}

We proceeded to extract the \chandra\ spectra in the same way as described in B16, using the {\sc ciao} (v4.7, CALDB v4.6.5) tool {\sc specextract}. For the point sources, spectra were extracted from elliptical regions drawn by eye to encompass the shape of the off-axis \chandra\ PSF. For X-1 we used an ellipse with a semi-major axis of 2--3\arcsec\ and a semi-minor axis of 1--2\arcsec. For X-2 the major and minor axes were 2\arcsec\ and 1\arcsec\ respectively. A small rectangular region close by was used for background subtraction. Figure \ref{fig_chandra_img} shows examples of the extraction regions used.

We used the {\sc ciao} tool {\sc pileup\_map} to give an indication of the level of pileup in each observation. The output, which is in counts per frame, ranges from 0.06 to 0.35 at maximum, which occurs at the position of X-1. These numbers correspond to pileup fractions of $<5$\% for $<0.1$ counts per frame to $>10$\% for $>0.2$ counts per frame. For three observations, obsIDs 17678, 18065 and 18067, X-1 has a pileup fraction of $>10$\%. We will discuss the potential effects of this on our results later in the paper.

For X-ray emission from M82 that does not come from X-1 or X-2, but that contributes to the emission seen by \nustar, such as from the fainter point sources and the diffuse emission, we extract spectra from a 49\arcsec\ radius circular region centered on X-1, but with the X-1 and X-2 regions as described above masked out. A larger background region external to the galaxy was extracted in order to assess the background coming from the Cosmic X-ray and particle backgrounds.

\subsection{NuSTAR}

The raw \nustar\ data were reduced using the {\sc nustardas} software package version 1.4.1 and CALDB version 20150316. The events were cleaned and filtered using the {\tt nupipeline} script with standard parameters. The {\tt nuproducts} task was used to generate the spectra and the corresponding response files. Spectra were extracted from a circular aperture of radius 49\arcsec\ centered on the peak of the emission. The background spectra were extracted from a circular region encompassing the same detector chip as the source,  with a radius of 118\arcsec, excluding the source extraction region and avoiding the wings of the PSF as much as possible. Data from both focal plane modules (FPMA and FPMB) were used for simultaneous fitting, without co-adding. 

\subsection{Swift}

\swift\ conducted monitoring of M82 with a typical cadence of a few days between 2012--2018. Since this monitoring ran contemporaneously with our \chandra\ and \nustar\ observations, the well-sampled lightcurve provides us with context for our study. A total of 113 observations, mostly consisting of obsIDs 00032503099--154 and 00092202001--051, have been made of the galaxy over the 2015--2016 period which we use here to calculate a long-term lightcurve. 

\begin{figure}
\begin{center}
\includegraphics[width=95mm]{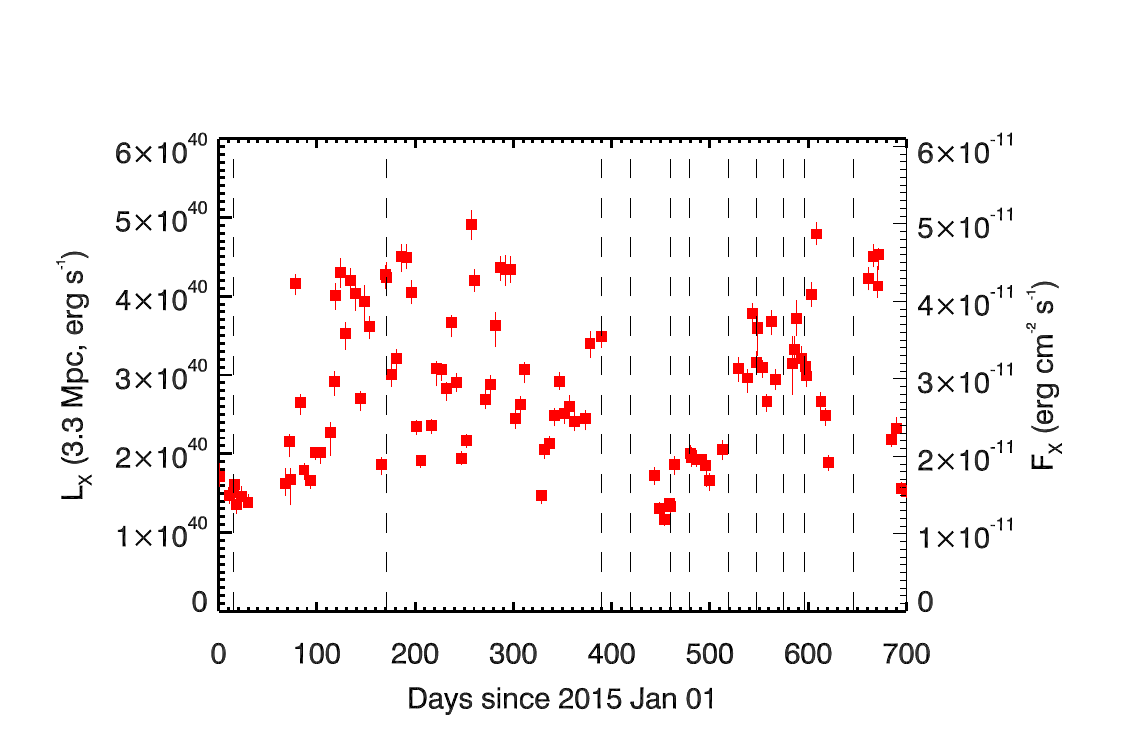}
\caption{0.5--8 keV lightcurve of the total observed X-ray flux from M82 using \swiftxrt. The highly variable X-ray emission is caused by X-1. The times of the simultaneous \chandra\ and \nustar\ observations we use here to study the spectral evolution of X-1 are shown with dashed lines.}
\label{fig_swift_ltcrv}
\end{center}
\end{figure}

We calculate the fluxes via spectral fitting. We use the {\sc heasoft} (v 6.16) tool {\sc xselect} to filter events from a 49\arcsec\ radius region centered on the ULXs and to extract the spectrum. This extraction region encloses all sources of X-ray emission in the galaxy. Background events were extracted from a nearby circular region of the same size. We group the spectra with a minimum of one count per bin using the {\sc heasoft} tool {\sc grppha}. We conduct spectral fitting in the range 0.2--10 keV. We fit the spectra with a simple power-law subjected to absorption intrinsic to M82 at $z=0.00067$ ({\tt zwabs*powerlaw} in {\sc xspec}) with the Cash statistic \citep{cash79} which uses a Poisson likelihood function and is hence most suitable for low numbers of counts per bin. From this model we calculate the observed flux in the 0.5--8 keV range, equivalent to the \chandra\ band. The lightcurve is presented in Figure \ref{fig_swift_ltcrv}.

The \swift\ lightcurve shows that our observations took place during a period of flaring activity from M82, which we found to be due to increased activity from X-1 (B16). The first \chandra/\nustar\ observations took place before the increase in activity, and the remaining observations tracked the activity over the next two years.

\section{Spectral analysis}
\label{sec_spec}

\begin{figure}
\begin{center}
\includegraphics[width=95mm]{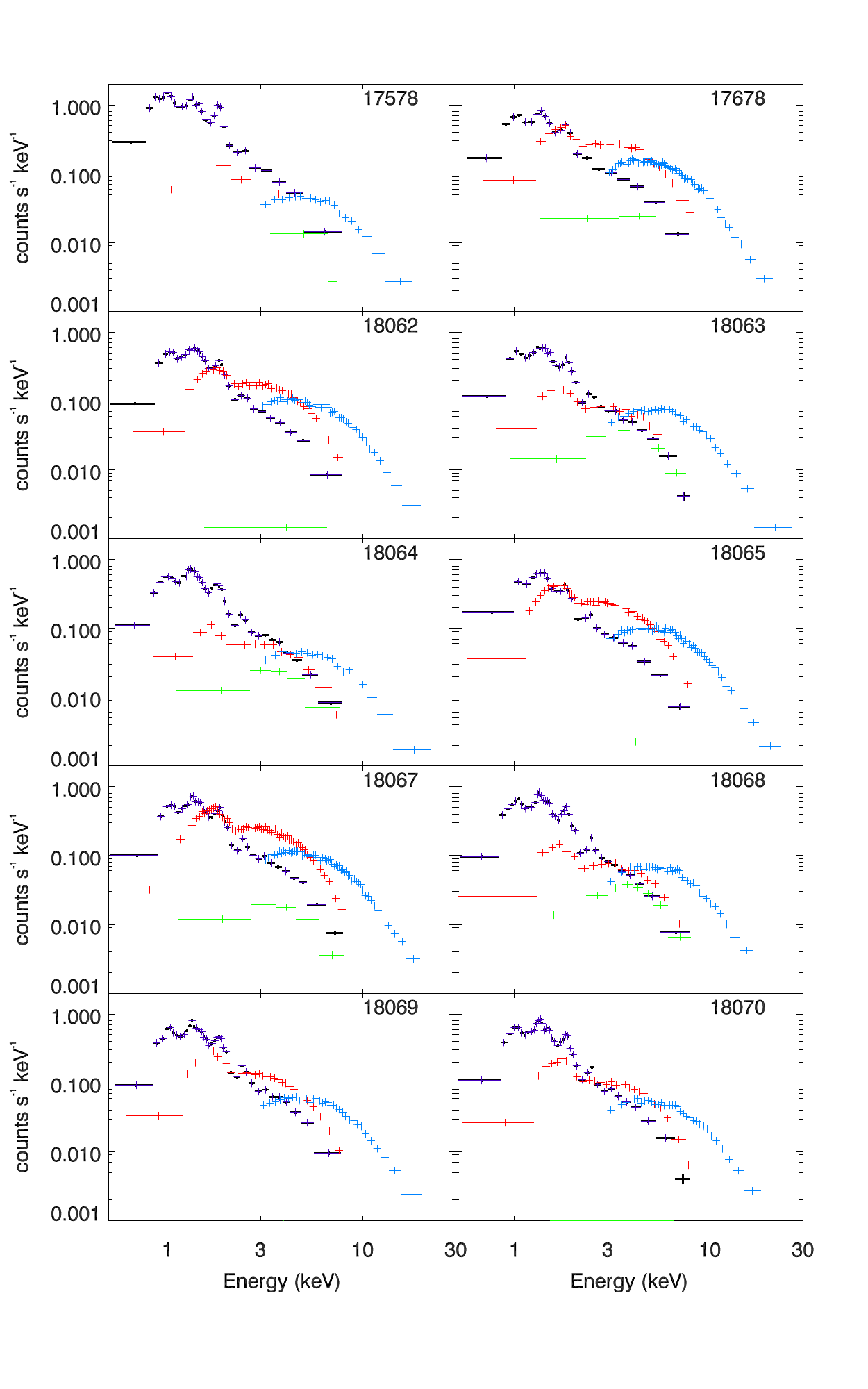}
\caption{\chandra\ spectra of X-1 (red), X-2 (green) and the diffuse emission (purple), and the \nustar\ FPMA spectra of all sources (blue) for each of the ten observational epochs listed in Table \ref{table_obsdat}. The spectra have been heavily rebinned for clarity.}
\label{fig_ld_spec}
\end{center}
\end{figure}

All \chandra\ and \nustar\ spectra were grouped with a minimum of 20 counts per bin using the {\sc heasoft} tool {\tt grppha}. Spectral fitting was carried out using {\sc xspec} v12.8.2 \citep{arnaud96} and the \chisq\ statistic was used for spectral fitting to background subtracted spectra. All uncertainties quoted are 90\%. We present the \chandra\ spectra of X-1, X-2 and the additional X-ray emission from M82, and the \nustar\ spectra of the entire galaxy for each of the 10 observational epochs in Figure \ref{fig_ld_spec}.

While the presence of X-2 in a bright state introduced some ambiguity to the results on the spectral analysis of X-1 in B16, in four observations here, \chandra\ obsIDs 18062, 18065, 18069 and 18070, X-2 was observed in a low state, allowing an unambiguous view of X-1 for the first time using \nustar. For these four observations, we neglect the emission from X-2 in our spectral fits,  whereas for the rest of the observations we included it.
 
 We proceed to conduct a spectral analysis for all 10 observations, where we fitted the spectra in the same way as described in B16. A cross-calibration constant was applied to each spectral data set to allow for absolute differences in normalization, and allowed to vary by $\pm10$\% \citep{madsen15}. For the diffuse emission from M82 we use a combination of three absorbed {\tt zwabs*apec} models with the temperatures and abundances fixed to the values found in B16. We allow the normalizations,  both with respect to B16 and to each other, to vary here to account for small differences in the detector responses. For the spectrum of X-1, we use the {\tt zwabs*diskpbb} model, where {\tt zwabs} is a redshifted neutral absorption component and {\tt diskpbb} is a model representing emission from a multicolour accretion disk, with a variable radial temperature profile. This model combination was found to best represent the emission from X-1 with regards to other disk models in B16. Additionally, since X-1 is a bright source, and despite measures taken to reduce pileup, the source still suffers from pileup. We account for this in spectral fitting using the {\tt pileup} convolution model, with frame time set to 0.4\,s \citep{davis01}. 

\begin{figure}
\begin{center}
\includegraphics[width=70mm]{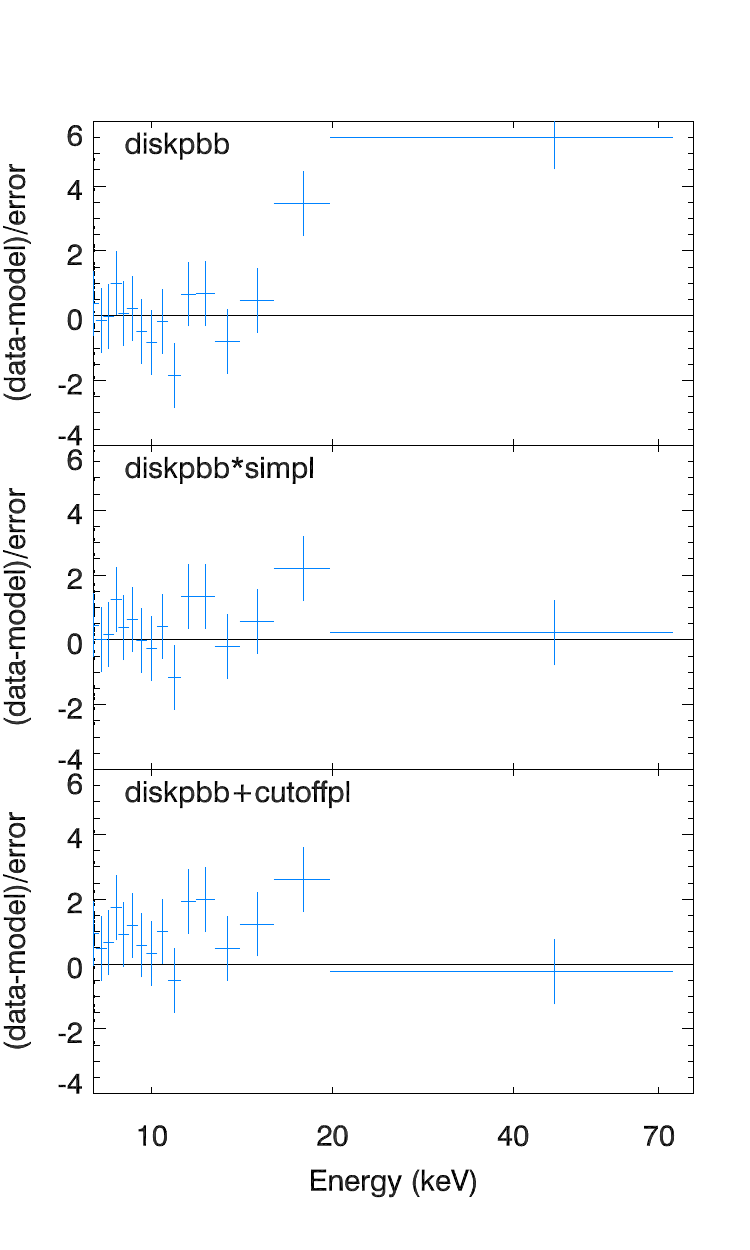}
\caption{Spectral residuals for a fit to the 18062 dataset on M82 X-1 where X-2 is off and a clear view of the source is seen. A prominent hard excess is seen with \nustar\ when fitted with a {\tt pileup*zwabs*diskpbb} model (top), which we account for with a {\tt simpl} model (middle) and {\tt cutoffpl} model (bottom).}
\label{fig_delc_spec}
\end{center}
\end{figure}

In the top panel of Figure \ref{fig_delc_spec} we show the residuals to this set of models which reveal strong residuals above 10 keV, especially in obsID 18062, which is reminiscent of the hard tail seen in the \nustar\ spectra of several other ULXs such as Holmberg II X-1 \citep[e.g.][]{walton15}, including those already identified as neutron star accretors such as NGC 7793 P13 \citep{walton17} and NGC 5907 ULX \citep{fuerst17,walton18c}. This was also seen in our analysis in B16, but since X-2 was bright during that observation, there was ambiguity regarding its origin. Here it is clear that it originates from X-1.

We test two models to fit this component, {\tt simpl}, which describes the power-law emission from the Comptonization of a disk spectrum \citep{steiner09}, and a {\tt cutoffpl} model used to model the pulsed emission from ULXPs \citep{brightman16,walton17,walton18c}. In \cite{walton18c}, where no pulsations from a source had been detected, as is the case for M82 X-1, $\Gamma$ was fixed at 0.5 for the {\tt cutoffpl} model which is the average for the pulsed emission from ULXPs. Therefore we do the same when using the {\tt cutoffpl} model and fix $\Gamma=0.5$. The energy of the cut off was allowed to vary. We find that the {\tt simpl} model provides a better description of the data, with \dchisq=70--200 for the addition of two free parameters from {\tt simpl} and \dchisq=3--90 for the addition of two free parameters from {\tt cutoffpl} as shown in the middle and bottom panels of Figure \ref{fig_delc_spec}. We do not find a significant improvement for the {\tt cutoffpl} model if we allow $\Gamma$ to vary. Therefore we use the {\tt simpl} model to fit the high energy emission from X-1 for the full dataset.

\begin{figure}
\begin{center}
\includegraphics[width=95mm]{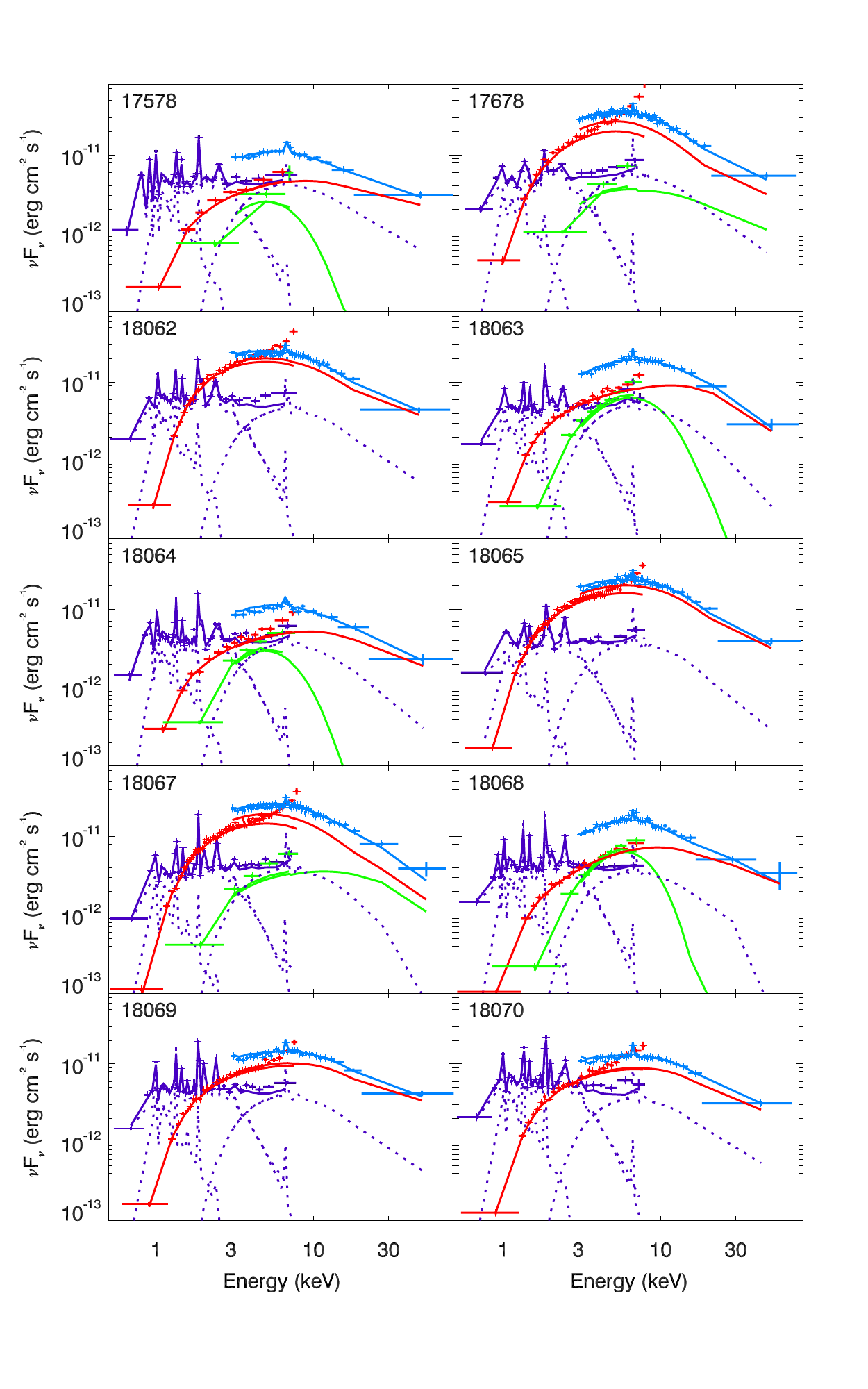}
\caption{\chandra\ spectra of X-1 (red), X-2 (green) and the diffuse emission (purple), and the \nustar\ FPMA spectra of all sources (blue) unfolded through the spectral responses with the assumed spectral models.}
\label{fig_eeuf_spec}
\end{center}
\end{figure}

We proceed to fit the \chandra\ and \nustar\ spectra of all 10 observational epochs with the {\tt pileup*zwabs*diskpbb*simpl} model combination to describe the emission from X-1. For the six observations where X-2 is bright, we model the emission from this source with an absorbed cut-off power-law model, {\tt zwabs*cutoffpl}, which was used in B16 to model the pulsed emission. Figure \ref{fig_eeuf_spec} presents the unfolded spectra with the different model components shown. Table \ref{table_specpar} presents the best-fit spectral parameters for X-1, and Table \ref{table_specpar2} presents the best-fit spectral parameters for X-2. 

In order to calculate the intrinsic luminosity of each source, we use the {\tt cflux} model component in {\tt xspec} placed {\it after} the absorption component with the normalization of the main model fixed. We calculate the intrinsic flux over the range 0.5--30 keV, and present these values with their uncertainties in Table \ref{table_specpar} and Table \ref{table_specpar2}. We calculate the intrinsic luminosities over these ranges assuming a distance to M82 of 3.3 Mpc \citep{foley14}.

\begin{table*}
\centering
\caption{Spectral fitting results for X-1}
\label{table_specpar}
\begin{center}
\begin{tabular}{cccccccccc}
\hline
\chandra\	& \nh\ & $T_{\rm in}$ 	& $p$	& log$_{10}$ norm	&  $\Gamma$	& \fscatt\ & \chisq/DoF	& \fx\ & \lx\ \\
 ObsID	& (10$^{22}$ \cmsq)	&  (keV)	& 	& 	&  & &	&(10$^{-11}$ \ergcms) & ($10^{40}$ \ergs)\\

  17578& 0.98$^{+ 0.15}_{- 0.33}$ & 2.8$^{+ 2.7}_{- 1.9}$ & 0.59$^{+ 0.13}_{- 0.03}$ &-2.6$^{+ 2.2}_{- 1.2}$ & 3.0$^{+ 2.0}_{- 2.0}$ & 0.98$^{+ 0.02}_{- 0.93}$ & 941.5/ 840& 0.97$^{+ 0.04}_{- 0.04}$ & 1.3$^{+ 0.1}_{- 0.1}$ \\
  17678& 1.35$^{+ 0.17}_{- 0.08}$ & 1.7$^{+ 0.4}_{- 0.4}$ & 0.64$^{+ 0.08}_{- 0.03}$ &-0.8$^{+ 0.4}_{- 0.4}$ & 3.7$^{+ 0.7}_{- 0.5}$ & 0.73$^{+ 0.27}_{- 0.40}$ &1984.5/1259& 4.26$^{+ 0.19}_{- 0.13}$ & 5.6$^{+ 0.3}_{- 0.2}$ \\
  18062& 1.81$^{+ 0.13}_{- 0.13}$ & 2.3$^{+ 0.8}_{- 0.7}$ & 0.52$^{+ 0.03}_{- 0.02}$ &-1.8$^{+ 0.6}_{- 0.5}$ & 3.3$^{+ 0.4}_{- 0.8}$ & 0.60$^{+ 0.40}_{- 0.47}$ &2070.3/1246& 3.73$^{+ 0.08}_{- 0.15}$ & 4.9$^{+ 0.1}_{- 0.2}$ \\
  18063& 1.25$^{+ 0.07}_{- 0.11}$ & 6.7$^{+ 0.6}_{- 3.1}$ & 0.57$^{+ 0.01}_{- 0.01}$ &-3.7$^{+ 0.7}_{- 0.2}$ & 4.5$^{+ 0.5}_{- 4.5}$ & 0.00$^{+ 0.26}_{- 0.00}$ &1445.8/1282& 1.82$^{+ 0.05}_{- 0.05}$ & 2.4$^{+ 0.1}_{- 0.1}$ \\
  18064& 1.09$^{+ 0.16}_{- 0.15}$ & 4.5$^{+ 1.7}_{- 2.3}$ & 0.58$^{+ 0.04}_{- 0.03}$ &-3.3$^{+ 0.9}_{- 0.6}$ & 2.7$^{+ 2.4}_{- 2.7}$ & 0.23$^{+ 0.77}_{- 0.23}$ &1486.0/1066& 1.04$^{+ 0.03}_{- 0.03}$ & 1.4$^{+ 0.0}_{- 0.0}$ \\
  18065& 1.25$^{+ 0.08}_{- 0.08}$ & 2.3$^{+ 0.7}_{- 0.3}$ & 0.57$^{+ 0.02}_{- 0.02}$ &-1.6$^{+ 0.2}_{- 0.4}$ & 3.7$^{+ 1.3}_{- 0.6}$ & 0.92$^{+ 0.08}_{- 0.60}$ &2069.2/1404& 4.21$^{+ 0.16}_{- 0.10}$ & 5.5$^{+ 0.2}_{- 0.1}$ \\
  18067& 1.38$^{+ 0.11}_{- 0.10}$ & 2.0$^{+ 0.5}_{- 0.6}$ & 0.56$^{+ 0.03}_{- 0.02}$ &-1.4$^{+ 0.5}_{- 0.2}$ & 3.5$^{+ 1.1}_{- 0.7}$ & 0.58$^{+ 0.42}_{- 0.36}$ &2410.8/1567& 3.10$^{+ 0.12}_{- 0.11}$ & 4.0$^{+ 0.2}_{- 0.1}$ \\
  18068& 0.94$^{+ 0.12}_{- 0.14}$ & 3.8$^{+ 0.9}_{- 1.7}$ & 0.64$^{+ 0.06}_{- 0.02}$ &-2.7$^{+ 1.0}_{- 0.4}$ & 2.4$^{+ 1.1}_{- 1.4}$ & 0.31$^{+ 0.58}_{- 0.24}$ &1502.7/1366& 0.94$^{+ 0.07}_{- 0.05}$ & 1.2$^{+ 0.1}_{- 0.1}$ \\
  18069& 1.25$^{+ 0.10}_{- 0.09}$ & 3.6$^{+ 1.3}_{- 1.0}$ & 0.56$^{+ 0.02}_{- 0.02}$ &-2.7$^{+ 0.5}_{- 0.3}$ & 2.7$^{+ 1.1}_{- 1.4}$ & 0.28$^{+ 0.33}_{- 0.26}$ &1529.0/1275& 2.31$^{+ 0.09}_{- 0.06}$ & 3.0$^{+ 0.1}_{- 0.1}$ \\
  18070& 1.30$^{+ 0.10}_{- 0.11}$ & 3.1$^{+ 0.6}_{- 0.5}$ & 0.56$^{+ 0.02}_{- 0.02}$ &-2.5$^{+ 0.3}_{- 0.1}$ & 4.0$^{+ 1.0}_{- 1.4}$ & 0.91$^{+ 0.09}_{- 0.86}$ &1598.2/1252& 2.02$^{+ 0.05}_{- 0.05}$ & 2.6$^{+ 0.1}_{- 0.1}$ \\

\hline
\end{tabular}
\tablecomments{Best-fit parameters for the {\tt diskpbb} model fitted to X-1. Fluxes and luminosities are given in the 0.5$-$30~keV range, and are corrected for absorption and pileup. Luminosities are calculated assuming a distance of 3.3 Mpc to M82.}

\end{center}
\end{table*}

\begin{table*}
\centering
\caption{Spectral fitting results for X-2}
\label{table_specpar2}
\begin{center}
\begin{tabular}{cccccccc}
\hline
\chandra\ 	& \nh\ 	& $\Gamma$	& $E_{\rm C}$  & log$_{10}$ norm	& \chisq/DoF	& \fx\ & \lx\ \\
 ObsID	& (10$^{22}$ \cmsq)	& 	&  (keV) & 	& 	&  (10$^{-11}$ \ergcms) &($10^{40}$ \ergs)\\

  17578& 1.7$^{+ 2.4}_{- 1.4}$ &-1.3$^{+ 3.4}_{- 3.6}$ & 1.5$^{+22}_{- 0.4}$ &-3.6$^{+ 0.9}_{- 0.5}$ & 899.9/ 840& 0.4$^{+ 0.3}_{- 0.1}$ & 0.5$^{+ 0.4}_{- 0.1}$ \\
  17678& 3.8$^{+ 1.2}_{- 1.1}$ & 1.7$^{+ 0.7}_{- 0.7}$ &15$^{+77}_{-16}$ &-2.7$^{+ 0.6}_{- 0.4}$ &1378.4/1259& 1.4$^{+ 0.5}_{- 0.3}$ & 1.8$^{+ 0.6}_{- 0.4}$ \\
  18063& 1.9$^{+ 0.4}_{- 0.4}$ &-1.2$^{+ 0.4}_{- 0.4}$ & 2.0$^{+ 0.4}_{- 0.3}$ &-3.6$^{+ 0.2}_{- 0.1}$ &1370.8/1282& 1.0$^{+ 0.1}_{- 0.1}$ & 1.2$^{+ 0.1}_{- 0.1}$ \\
  18064& 2.3$^{+ 1.3}_{- 1.0}$ &-1.3$^{+ 1.3}_{- 1.6}$ & 1.4$^{+ 0.9}_{- 0.4}$ &-3.4$^{+ 0.4}_{- 0.4}$ &1233.9/1066& 0.5$^{+ 0.2}_{- 0.1}$ & 0.6$^{+ 0.3}_{- 0.1}$ \\
  18067& 3.0$^{+ 0.7}_{- 0.6}$ & 1.3$^{+ 0.5}_{- 0.6}$ &16$^{+25}_{-10}$ &-3.0$^{+ 0.3}_{- 0.2}$ &1725.4/1567& 1.0$^{+ 0.1}_{- 0.2}$ & 1.3$^{+ 0.2}_{- 0.2}$ \\
  18068& 0.8$^{+ 0.2}_{- 0.8}$ &-2.9$^{+ 6.9}_{- 2.0}$ & 1.2$^{+ 0.0}_{- 0.1}$ &-4.0$^{+ 0.0}_{- 0.0}$ &1478.3/1366& 0.7$^{+ 0.0}_{- 0.0}$ & 1.0$^{+ 0.1}_{- 0.0}$ \\

\hline
\end{tabular}
\tablecomments{Best-fit parameters for the {\tt cutoffpl} model fitted to X-2. Fluxes and luminosities are given in the 0.5$-$30~keV range, and are corrected for absorption. Luminosities are calculated assuming a distance of 3.3 Mpc to M82.}

\end{center}
\end{table*}

\section{Spectral evolution of M82 X-1}
\label{sec_specevol}

Our main goal is to explore the spectral evolution of M82 X-1. For each pair of spectral parameters, we compute Spearman's rank correlation, using the {\sc idl} tool {\tt r\_correlate.pro}, to assess the presence of any correlation and its significance. This assesses how well the relationship between two variables can be described using a monotonic function. The results from this test, being the rank correlation coefficient, $\rho$, and the two-sided significance of its deviation from zero, $p$, are presented in Table \ref{table_speccor}.

The significance is a value in the interval [0.0, 1.0] and a small value indicates a significant correlation. The commonly used threshold to judge that a correlation is significant is $p<0.05$. We find that this criterion is met for the following pairs of parameters: \lx\ and \nh, \lx\ and $T_{\rm in}$, \lx\ and log$_{\rm 10}$norm, $T_{\rm in}$ and log$_{\rm 10}$norm, \lx\ and \fscatt, $T_{\rm in}$ and \fscatt, and $\Gamma$ and \fscatt. 

However, for many parameters, especially those of the {\tt simpl} model, the uncertainties are large, which the correlation test does not account for. First, in order to account for this, we conduct 1000 Monte-Carlo simulations where for each pair of parameters we randomly draw a value from the 90\% confidence interval. We calculate from how many of the 1000 simulations do we find a correlation that has a $p$ value which is less than 0.05. For all correlations involving parameters of the {\tt simpl} model, we find that in less than 25\% of the simulations, a $p$ value less than 0.05 is recovered. Therefore we do not have confidence in these correlations being real due to the large uncertainties in the parameters.

\begin{figure*}
\begin{center}
\includegraphics[width=180mm]{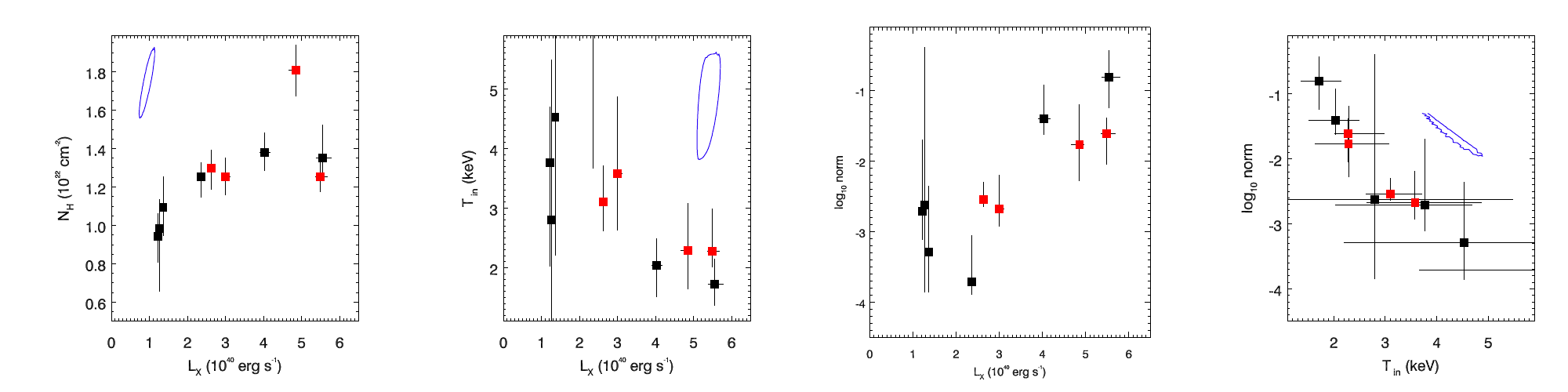}
\caption{The relationship between the spectral parameters of the {\tt zwabs*simpl*diskpbb} model for X-1 where a significant correlation was found. Red data points indicate observations where X-2 is at low fluxes and thus the view of the emission from X-1 is unambiguous. We also overplot the 3-$\sigma$ \chisq\ contour in blue, shifted for clarify, from \chandra\ obsID 18069 to demonstrate if the correlation is driven by degeneracy between the parameters, which appears to be the case for $T_{\rm in}$ and log$_{\rm 10}$norm.}
\label{fig_specpar}
\end{center}
\end{figure*}

Furthermore, for the correlations where the uncertainties in the parameters are smaller, some of these correlations can be expected due to degeneracies between parameters. In order to determine if these correlations are driven by degeneracies, we explore the two-dimensional \chisq\ space around the best fit using the {\sc xspec} command {\tt steppar}, and overplot the 3-$\sigma$ contours on Figure \ref{fig_specpar}. We do this only for one observation, \chandra\ obsID 18069, since it is computationally expensive, but this should be adequate to reveal any spectral degeneracies. This dataset is a typical observation where X-2 is at low fluxes,  pileup for X-1 is low, and the measured parameters are in the middle of the distributions. 

The contours show that there is slight degeneracy between \lx\ and \nh, but that does not appear to be strong enough to induce the correlation. There is no apparent degeneracy between \lx\ and $T_{\rm in}$. A correlation between \lx\ and log$_{\rm 10}$norm is expected and a clear degeneracy between $T_{\rm in}$ and log$_{\rm 10}$norm is seen. 

Since the background is higher in the \nustar\ data than in the \chandra\ data, and signal to noise lower, we investigated whether binning the \nustar\ spectra with more counts would affect our results. We grouped the \nustar\ spectra with a minimum of 60 counts per bin, rather than 20, and refit the spectra. We looked at the temperature of the {\tt diskpbb} component, which is most sensitive to the NuSTAR data and the one of the key parameters involved in our results. We found the average difference in temperature between the stronger binning and the weaker one to be -0.02 keV, which is much smaller than the typical uncertainty on the parameter. We therefore conclude that the spectral binning does not affect our results.

Taking into account the uncertainties in parameters and degeneracies between them, we can only say with confidence that there is a correlation and therefore physical link between the neutral column density and the X-ray luminosity, and the the inner disk temperature and X-ray luminosity.

We test the apparent dependence of \nh\ on \lx\ further by fixing \nh\ at the approximate mean value of $1.3\times10^{22}$ \cmsq\ in all fits and note the difference in \chisq. In all cases \chisq\ is worse or the same as when \nh\ is a free parameter. For \chandra\ obsID 18062 \lx\ is high and the \nh\ measured is particularly high at $1.8\times10^{22}$ \cmsq. When \nh\ is fixed at the mean value, the difference in \chisq\ is 80. For obsID 18068 \lx\ is low and \nh$=9\times10^{21}$ \cmsq. When fixed at $1.3\times10^{22}$ \cmsq, the \chisq\ increases by 150, supporting the finding that \nh\ does indeed depend on \lx.

The anti-correlation between \lx\ and $T_{\rm in}$ that we find here is in contrast with the correlation found between these two parameters for M82 X-1 by \cite{feng10}, where the apparent \lx${\propto}T^4$ relationship lead the authors to conclude that the source was observed in the thermal dominant state. In order to determine the exponent on the relationship we have found, we take the logarithm of both \lx\ and $T_{\rm in}$, such that log$_{10}$\lx$=\alpha$log$_{10}T_{\rm in}+\beta$ and conduct a linear regression analysis. We use the {\sc idl} tool {\tt linmix\_err.pro} which takes into account uncertainties in both parameters. We find that log$_{10}$\lx$=(-1.47\pm 1.00)$log$_{10}T_{\rm in}+(41.16\pm 0.47)$, where the uncertainties are $1\sigma$. Therefore the exponent is $-1.47\pm 1.00$, which is a $>5\sigma$ difference from the results from F10, where the exponent was implied to be 4, although the uncertainty in this parameter was not listed. If we run the same linear regression analysis on the results presented in F10, we find that the relationship is essentially unconstrained.

The key differences between our analysis and that of F10 are that they did not have accompanying \nustar\ data, and so their analysis was restricted to the narrow bandpass of 0.7--7 keV. The spectral models used are also different, whereby F10 used the {\tt diskbb} model, which is related to {\tt diskpbb} that we use when $p$ is fixed at 0.75. However, as we showed in B16, and confirm here, the data are inconsistent with the $p=0.75$ that describes a geometrically thin accretion disk. Furthermore, without \nustar\ data, the high energy excess that we detect and fit with {\tt simpl} was not observable by F10. Finally, the uncertainties related to degeneracies in the {\tt pileup} model are reduced when data from an instrument without pileup such as \nustar\ is used.

We investigate what effect pileup has on our results by exploring the dependence of \lx, \nh\ and $T_{\rm in}$ on the {\tt pileup} model parameter, $\alpha$. We find that none of these parameters show any dependence on $\alpha$, leading us to conclude that this model component does not drive our results. Furthermore, as mentioned earlier, for three observations, odsID 17678, 18065 and 18067, the pileup fraction for X-1 is greater than 10\%, and therefore the {\tt pileup} model used in the spectral fitting may not be able to reliably account for this effect. We test the dependence of our results on these observations by removing them from our analysis. In doing so, we still find a significant correlation between  \lx\ and \nh, and indeed the significance increases, but for \lx\ and $T_{\rm in}$ the correlation is no longer significant. {\bf A fit with a linear relationship reveals log$_{10}$\lx$=(-1.23\pm 2.29)$log$_{10}T_{\rm in}+(41.03\pm 1.28)$ and therefore we can only rule out a \lx${\propto}T^4$ dependency at $\sim2\sigma$}.

{\bf Finally, we note that some of our spectral fits have large \chisq\ relative to the number of degrees of freedom (DoF), indicating an unacceptable fit. This is likely due to the fits being quite complex, with many DoFs and data sets being fitted simultaneously. However, we note the presence of a potential excess of counts in the \nustar\ data at 3--4 keV that has been found in bright X-ray binaries and is a known calibration issue. We find that if we ignore data from \nustar\ below 5 keV, the fits improve and the result of this is to systematically reduce the temperature of the {\tt diskpbb} component. This reduction is consistent with the uncertainties on this component, however, and does not alter our result that there is an anti-correlation between \lx\ and $T_{\rm in}$ since the effect is systematic across all observations.}

These findings therefore rule out a thermal state for sub-Eddington accretion and therefore do not support M82 X-1 as an IMBH candidate.

\begin{table}
\centering
\caption{Spearman's rank correlation results for X-1}
\label{table_speccor}
\begin{center}
\begin{tabular}{c c c c c c c c}
\hline
     & \lx & \nh & T$_{\rm in}$ & p & log$_{\rm 10}$norm & $\Gamma$ \\
\multirow{2}{*}{\nh} & 0.76& & & & & \\
&  0.0111& & & & & \\
\multirow{2}{*}{T$_{\rm in}$} &-0.76&-0.56& & & & \\
&   0.011&   0.090& & & & \\
\multirow{2}{*}{p} &-0.33&-0.62& 0.03& & & \\
&   0.347&   0.054&   0.934& & & \\
\multirow{2}{*}{log$_{\rm 10}$norm} & 0.79& 0.62&-0.99&-0.10& & \\
&   0.006&   0.054&   0.000&   0.777& & \\
\multirow{2}{*}{$\Gamma$} & 0.48& 0.58&-0.22&-0.36& 0.28& \\
&   0.162&   0.082&   0.533&   0.310&   0.425& \\
\multirow{2}{*}{\fscatt} & 0.28& 0.09&-0.67& 0.05& 0.65& 0.21\\
&   0.425&   0.803&   0.033&   0.881&   0.043&   0.043\\

\hline
\end{tabular}
\tablecomments{For each pair of parameters we list the rank correlation coefficient (top) and the two-sided significance of its deviation from zero (bottom).}
\end{center}
\end{table}

\section{New insights into M82 X-2}
\label{sec_x2}

\begin{figure*}
\begin{center}
\includegraphics[width=180mm]{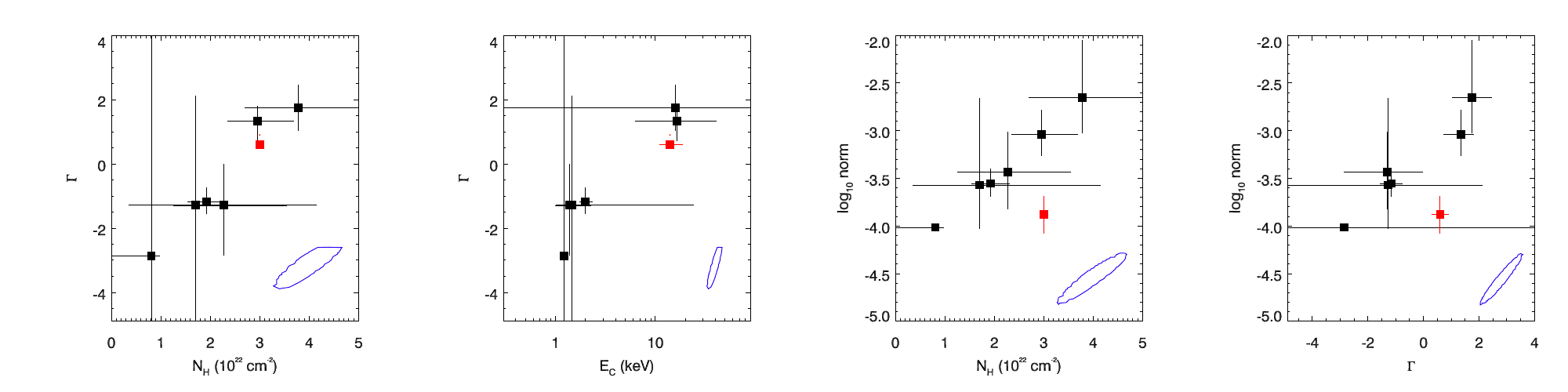}
\caption{The relationship between the spectral parameters of the {\tt zwabs*cutoffpl} model for X-2 where evidence for a correlation has been found. Red data points indicate the best-fit parameters of the pulsed emission. We also overplot the 3-$\sigma$ \chisq\ contour in blue, shifted for clarify, from \chandra\ obsID 18063 to demonstrate if the correlation is driven by degeneracy between the parameters, which appears to be the case for all parameter pairs}
\label{fig_specpar2}
\end{center}
\end{figure*}

We conduct the same analysis for X-1 on X-2, where the spectral parameters are plotted against each other in Figure \ref{fig_specpar2} and the correlation analysis is shown in Table \ref{table_speccor2}. Here our analysis suggests correlations between \nh\ and $\Gamma$, $E_{\rm C}$ and $\Gamma$, \nh\ and log$_{\rm 10}$norm, and  $\Gamma$ and log$_{\rm 10}$norm. However, the \chisq\ contours show that it is likely that strong degeneracies between these parameters drive these apparent correlations (Fig \ref{fig_specpar2}). We therefore don't find evidence for any significant spectral evolution in X-2.

While the emission from X-2 is difficult to disentangle from the other sources of emission in M82, we were able to isolate the pulsed emission in the \nustar\ band from this source in \cite{brightman16}. We found that the pulsed emission is best fit by a power-law with a high-energy cut-off, where $\Gamma=0.6\pm0.3$ and $E_{\rm C}=14^{+5}_{-3}$ keV. In Figure \ref{fig_specpar2} we show the parameters of the pulsed emission as a separate data point for comparison. We see that the values for $\Gamma$ and $E_{\rm C}$ from our broadband fits are consistent with the pulsed emission when X-2 is at its highest luminosities, \lx$>10^{40}$ \ergs, indicating that at these times the pulsations are most likely to be detected. \cite{bachetti19} have recently detected pulsations again from a \nustar\ observation taken on 2016-09-10, which unfortunately did not have any simultaneous \chandra\ observations, and so we did not include it in our analysis here.

\begin{table}
\centering
\caption{Spearman's rank correlation results for X-2}
\label{table_speccor2}
\begin{center}
\begin{tabular}{c c c c c c c c}
\hline
     & \lx & \nh & $E_{\rm C}$ & $\Gamma$ \\
\multirow{2}{*}{\nh} & 0.71& & &\\
&   0.111& & &\\
\multirow{2}{*}{E$_{\rm C}$} & 0.71& 0.77& & \\
&   0.111&   0.072& & \\
\multirow{2}{*}{$\Gamma$} & 0.77& 0.83& 0.94& \\
&   0.072&   0.042&   0.005& \\
\multirow{2}{*}{log$_{\rm 10}$norm} & 0.71& 1.00& 0.77& 0.83\\
&   0.111&   0.000&   0.072&   0.042\\

\hline
\end{tabular}
\tablecomments{For each pair of parameters we list the rank correlation coefficient (top) and the two-sided significance of its deviation from zero (bottom).}
\end{center}
\end{table}

\section{Discussion and Implications}

\subsection{M82 X-1 as an intermediate-mass black hole candidate}

M82 X-1 has been claimed to be an intermediate-mass black hole candidate based on its high X-ray luminosity, twin QPOs, and \lx${\propto}T^4$ scaling, all of which put the mass of the black hole at $\sim$10$^2$ \msol. Here we find that we can rule out a \lx${\propto}T^4$ scaling. This scaling relationship was expected from a standard accretion disk which exists at moderate accretion rates, and allows an estimate of the black hole mass from measurements of the inner edge of the accretion disk. Without these pieces of evidence, the status of M82 X-1 as an IMBH accretor is less certain.

\subsection{M82 X-1 as a super-Eddington accreting stellar-remnant}

The X-ray properties of M82 X-1 may be explained by it harboring a super-Eddington accreting stellar remnant black hole or neutron star. The spectral shape of M82 X-1, consisting of a broadened disk with a high energy tail is very similar to all other ULXs with high-quality broadband spectral data from \nustar\ \citep[e.g.][]{walton18c}. This sample includes the known neutron star accretors NGC~5907~ULX1 and NGC~7793~P13, and at first glance their spectral shapes are not dissimilar from the rest of the sample. This implies that these ULXs, including M82 X-1, are also super-Eddington accretors, although it is still not known whether they are powered by neutron stars or black holes.

One popular model to explain the spectral evolution of ULXs is that they are stellar-remnant black holes accreting at super-Eddington rates. In this model a powerful wind is radiatively driven from the accretion disk \citep[e.g.][]{poutanen07}, as recently revealed through the detection of highly ionized material in the high resolution X-ray spectra of NGC 1313 X-1 \citep{pinto16} among others \citep{pinto17,kosec18} and the detection of blueshifted iron-K absorption \citep{walton16}. Regarding the link between \nh\ and \lx\ as seen in M82 X-1, if this source is a stellar-mass black hole accreting at super-Eddington rates, as the mass accretion rate increases an increase in the X-ray luminosity follows, which drives further outflow of material from the system. Depending on the line of sight, this can cause an increase in the line of sight absorption. However, this material is expected to be highly ionized at small radii. \cite{middleton15b} also found for NGC 1313 X-1 that the neutral column density anti-correlates with spectral hardness, suggesting that at large radii, there is a cool, neutral component of the outflow, where the column density is linked to mass loss via increase mass accretion rate, as predicted by \cite{poutanen07}.

A negative relationship between \lx\ and $T_{\rm in}$ has been observed in other ULXs. For example \cite{luangtip16} studied the spectral evolution of Holmberg IX X-1, finding that the peak of the spectrum decreases with luminosity, suggesting an anti-correlation between \lx\ and $T$ (see also \cite{walton17a}). \cite{kajava09} also explored the relationship between \lx\ and $T$ in a sample of ULXs with \xmm\ and \chandra\ observations. While they found that for several sources fitted with a multicolor disk model follow the \lx${\propto}T^4$ scaling, sources with a disk plus power-law spectral shape show a negative \lx${\propto}T^-3.5$ scaling, similar to what we find here, and predicted by \cite{poutanen07}.

\section{Summary and Conclusions}
\label{sec_conc}

We have conducted a comprehensive investigation into the spectral evolution of the ultraluminous X-ray sources M82 X-1 and X-2 using ten simultaneous \chandra\ and \nustar\ observations. The \chandra\ data allowed us to spatially resolve the sources, separated by only 5\arcsec, while the \nustar\ data allowed a broadband X-ray spectral analysis. We found that for X-1, the luminosity of the disk scales with the inner temperature as $L{\propto}T^{-3/2}$, which is contrary to previous findings of a $L{\propto}T^{4}$ scaling that supported a standard accretion disk powering the system. We furthermore find evidence that the neutral column density of the material in the line of sight increases with \lx, perhaps due to an increased mass outflow with accretion rate. For X-2, we do not find any significant spectral evolution, but we can constrain the spectral parameters, and find the broadband emission to be consistent with the pulsed emission at the highest X-ray luminosities.

\section{Acknowledgements}

This research has made use of data obtained with \nustar, a project led by the California Institute of Technology, managed by the Jet Propulsion Laboratory, and funded by NASA. We thank the \nustar\ Operations, Software and Calibration teams for support with the execution and analysis of these observations. This research has made use of the \nustar\ Data Analysis Software (NuSTARDAS) jointly developed by the ASI Science Data Center (ASDC, Italy) and the California Institute of Technology (USA). Support for this work was provided by the National Aeronautics and Space Administration through \chandra\ Award Number GO6-17080X issued by the \chandra\ X-ray Center, which is operated by the Smithsonian Astrophysical Observatory for and on behalf of the National Aeronautics Space Administration under contract NAS8-03060. This research has made use of software provided by the Chandra X-ray Center (CXC) in the application package {\sc ciao}. We also acknowledge the use of public data from the {\it Swift} data archive. DJW acknowledges support from an STFC Ernest Rutherford Fellowship.

{\it Facilities:} \facility{\chandra\ (ACIS), \nustar, \swift\ (XRT)}


\end{document}